\documentclass[aps,pra,twocolumn,showpacs]{revtex4}
\usepackage{graphicx}
\usepackage{amsmath}
\usepackage{latexsym}
\usepackage{amsfonts}
\usepackage{amssymb}
\usepackage{array}
\usepackage{epsfig}
\newcommand{\bra}[1]{\left\langle{#1}\right\vert}
\newcommand{\ket}[1]{\left\vert{#1}\right\rangle}

\newcommand{\ketbra}[2]{|#1\rangle \langle#2|}

\newcommand{\ncd}{\newcommand}
\newcommand{\be}{\begin{equation}}
\newcommand{\ee}{\end{equation}}
\newcommand{\ba}{\begin{array}}
\newcommand{\ea}{\end{array}}
\newcommand{\bqa}{\begin{eqnarray}}
\newcommand{\eqa}{\end{eqnarray}}

\ncd{\SABC}{S^{ABC}}
\ncd{\Sab}{S^{ab}}
\ncd{\Sbc}{S^{bc}}
\ncd{\Sba}{S^{ba}}
\ncd{\csk}{{|\phi_{\{\kappa\} }
\rangle}_{\cal{C}}}
\ncd{\nbgh}{\text{nghb}}
\ncd{\QC}{$\mbox{QC}_{\cal{C}}$}

\begin{document}
\title{One-way quantum computing in a decoherence-free subspace}
\author{M. S. Tame}
\author{M. Paternostro}
\author{M. S. Kim}
\affiliation{School of Mathematics and Physics, The Queen's University, Belfast BT7 1NN, UK}
\date{\today}
\begin{abstract}
We introduce a novel scheme for one-way quantum computing (QC) based on the use of information encoded qubits in an effective cluster state resource. 
With the correct encoding structure, we show that it is possible to protect the entangled resource from phase damping decoherence, where the effective cluster state can be described as residing in a Decoherence-Free Subspace (DFS) of its supporting quantum system. One-way QC then requires either single or two-qubit adaptive measurements. As an example where this proposal can be realized, we describe an optical lattice setup where the scheme provides robust quantum information processing. We also outline how one can adapt the model to provide protection from other types of decoherence.
\end{abstract}

\maketitle
Quantum computing (QC) offers a huge advantage over its classical counterpart in terms of computational speedup of tasks such as database searching and number factorization~\cite{GrovShor}.  
These applications are expected to pave the way for the realization of a vast range of classically prohibitive computational tasks in both science and industry. 
An exciting new approach known as the one-way QC model~\cite{RBH} has attracted considerable interest from the theoretical~\cite{Hein} and experimental quantum information community~\cite{Wal,pan}. The basic ingredients of this computational model and the possibility of active feed-forwardability allowing fast gate performances have very recently been experimentally demonstrated~\cite{Wal}. In general the model is based on adaptive measurements in multipartite entangled resources known as graph states~\cite{HEB} which have been experimentally produced for the case of up to six qubits~\cite{pan}. It is a promising candidate for the implementation of quantum computing 
in physical systems where highly entangled resources can be generated in a massively parallelized fashion.

A particular class of graph states, known as cluster states, have proved to be crucial in one-way QC, as they form universal resources for QC based on single qubit adaptive measurements. However, the accuracy of protocols using cluster states is affected by sources of environmental decoherence and imperfections in the supporting quantum system~\cite{Tame1}. It is therefore desirable to design effective schemes to protect the quality of the entangled resources and the encoded information within. Quantum error-correction (QEC)~\cite{QEC} and the use of decoherence-free subspaces (DFS)~\cite{DFS} are two well-known methods that offer protection against loss of information from a supporting quantum system to the environment. The former requires a considerable overhead in system resources largely due to redundancy of the encoded information, while the latter requires a careful understanding of symmetries in the system-environment dynamics. The role of QEC in one-way QC has been studied previously \cite{QECcluster}, here we change perspective and discuss the application of DFS as a novel method for protecting quantum information during the performance of one-way QC. This approach requires significantly less physical qubits and adaptive measurements than a scheme based on QEC and puts our proposal closer to experimental implementation in far simpler physical setups. 
We first introduce, in Sec.~\ref{model}, a model for a quantum system that supports a multipartite entangled resource constituting an effective cluster state, invariant under random phase errors induced from scattering type decoherence in the system-environment dynamics. We then show how one-way QC can be carried out on this entangled resource with single or two-qubit adaptive measurements. In order to provide an operative way to evaluate the resilience to noise provided by the protection of the register by using a DFS, we outline a {\it quantum process tomography} technique that can easily be adapted to various experimental setups~\cite{imoto}. We quantitatively analyze the case of information transfer across a linear cluster state whose physical qubits are affected by phase damping decoherence and show the superiority of the DFS encoding. In Sec.~\ref{realization}, we provide a description of an optical lattice setup, where the required resource can be generated with cold controlled collisions and the measurements performed via Raman transitions and fluorescence techniques. Finally, Sec.~\ref{summary} summarizes our results and includes a brief outline of how our scheme can be adapted to provide protection from other forms of collective decoherence.

\section{The Model} 
\label{model}

\begin{figure}[t]
\centerline{
\psfig{figure=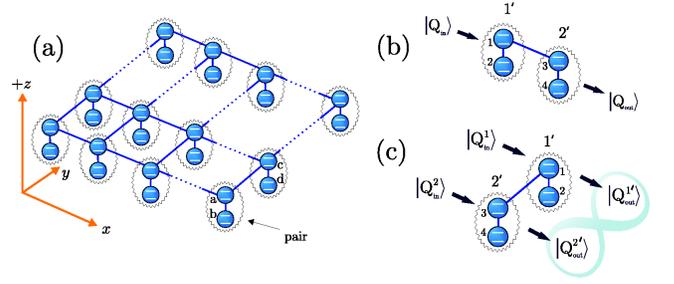,width=8.5cm}}
\caption{{\bf (a)}: The effective two-dimensional cluster state layout with each pair of physical qubits representing an encoded effective qubit. The qubits belonging to each pair couple to the environment in the same way, as described by Eq. (\ref{hamil}). {\bf (b)}: Schematics for information propagation of a logical qubit $\ket{Q_{\rm in}}$. {\bf (c)}: Schematics for the simulation of a gate operation on two logical qubits $\ket{Q^1_{\rm in}}$ and $\ket{Q^2_{\rm in}}$. The body of the paper provides a detailed account of the procedures to follow in {\bf (b)} and {\bf (c)}.}
\label{lattice}
\end{figure}

We consider a set of qubits occupying the sites of a lattice structure ${\cal C}$ as shown in Fig.~\ref{lattice} {\bf (a)}. Each pair of qubits is prepared in the singlet state 
\begin{equation}
\ket{\psi^-}_{ab}=\frac{1}{\sqrt{2}}(\ket{01}-\ket{10})_{ab}
\end{equation}
with $\{ \ket{0},\ket{1}\}$ the single-qubit basis. In what follows, each first (second) {\it pedex} labels a qubit belonging to the top (bottom) qubit-layer with respect to the positive $z$-axis (see Fig.~\ref{lattice} {\bf (a)}). The top qubits $a$ and $c$ of two neighbouring pairs are connected via the controlled-$\sigma_z$ operation  
\begin{equation}
S^{ac}=|0 \rangle_a \langle 0| \otimes {\openone}_{c} + |1 \rangle_a \langle 1| \otimes \sigma_{z,c}
\end{equation}
with $\sigma_{l,i}~(l=x,y,z)$ the $l$-Pauli matrix applied to qubit $i$. In order to generate this entanglement structure, one initially sets the top and bottom qubits $a,b$ to the state $\ket{-}$, where $\ket{\pm}=(1/\sqrt{2})(\ket{0}\pm \ket{1})$, resulting in the total state $\otimes_{a,b\in {\cal C}}\ket{-,-}_{ab}$. The transformation 
\begin{equation}
{\cal S}_{\|}^{\cal C}=\prod_{a,b\in {\cal C}|a,b \in \gamma_{\|}}S^{ab}
\end{equation}
is then applied to the qubits
along the $z$-axis, where $\gamma_{\|}=(0,0,1)^{T}$. This is followed by the operation $\prod_{a,b\in {\cal C}|a,b \in \gamma_{\|}}\openone_a \otimes H_b$, where $H_i$ is the Hadamard gate applied to qubit $i$, resulting in the state 
\begin{equation}
\bigotimes_{a,b\in {\cal C}|a,b\in \gamma_{\|}}\ket{\psi^-}_{ab}.
\end{equation}
The next step is the application of the transformation ${\cal S}_{=}^{\cal C}=\prod_{a,c\in {\cal C}|a,c\in \gamma_{=}}S^{ac}$ to qubits belonging to the top layer of the lattice, where $\gamma_{=}=\{(1,0,0)^{T},(0,1,0)^{T}\}$. We now consider the encoding $\{\ket{0_E}_{a'}:=\ket{01}_{ab}, \ket{1_E}_{a'}:=-\ket{10}_{ab} \}$, where each pair of physical qubits embodies an {\it effective} qubit $a'$ in a single-layer lattice ${\cal C'}$. The state generated in this way corresponds to a standard cluster state $\ket{\phi_{\{ \kappa \}}}_{\cal C'}$ with eigenvalue set $\{ \kappa \}$ containing $\kappa_{a'}=0,~(\forall a'\in {\cal C'})$~\cite{RBH}, which we denote, for ease of notation, as $\ket{\phi_{}}_{\cal C'}$. From now on, the top and bottom physical qubits encompassed in $a'$ will be labeled as $a'_1$ and $a'_2$ respectively. 

The physical assumption we make here concerning the noise affecting the prepared computational register is that while qubits in the $x$-$y$ plane across the lattice structure are at a fixed distance from each other, the qubits along the $z$-axis are closer together (see Fig. \ref{lattice} {\bf (a)}) such that each qubit in a pair couples to the environment in the same way. This means that the environment cannot distinguish the qubits and one can write the Hamiltonian for the paired-qubit system and the environment as \cite{DFS}
\be
\label{hamil}
H=E_{0}\otimes \openone +E_x \otimes J_x+E_y \otimes J_y+E_z \otimes J_z,
\ee
where $E_0$, $E_x$, $E_y$ and $E_z$ are the operators of the environment and we have $J_x=(1/2)\sum_{i=1}^{2}\sigma_{a'_i,x}$, $J_y=(1/2)\sum_{i=1}^{2}\sigma_{a'_i,y}$ and $J_z=(1/2)\sum_{i=1}^{2}\sigma_{a'_i,z}$. The hamiltonian in Eq. (\ref{hamil}) describes well the physical situation when both qubits are very close together when compared to the environment's coherence length \cite{DFS}, where a Markov approximation is implicit in the description. This is a reasonable assumption in many physical situations, one of which will be treated in detail in Sec.~\ref{realization}. The qubits affected by the collective type of noise described in Eq. (\ref{hamil}) are depicted by jagged surroundings in Fig.~\ref{lattice}. The encoded qubit state $\ket{+}_{a'}$ (equivalent to $\ket{\psi^{-}}_{a'_1a'_2}$), before entanglement generation on the top layer, is invariant under environment-induced phase shifts (associated with the final term in Eq. (\ref{hamil})) on the physical qubits, $\ket{j}\to e^{i \phi_j}\ket{j}~(j=0,1)$. As any random phase shifts of this form commute with the operations $S^{ac}$ on the top layer producing the encoded cluster state, the final state $\ket{\phi_{}}_{\cal C'}$ is unaffected by such an environment also. The {\it dual-rail} encoding we have used is well-known in providing robust protection against phase damping decoherence~\cite{DFS}. The combination of this encoding and the entangling operations we describe put the encoded cluster state $\ket{\phi_{}}_{\cal C'}$ in a DFS for the phase damping class of noise considered here, {\it i.e.} to describe the dynamics we set $E_x=E_y=0$ in Eq. (\ref{hamil}). The possibility of encoding within such a DFS is important in many physical setups where random phase fluctuations are the dominant source of decoherence. For example, in optical lattices and ion-traps this decoherence mechanism is caused by an environment at non-zero temperature exciting the motional states of the atoms that embody the physical qubits~\cite{Jak2, Dav1, Dav2, Dav3}.

In order to understand how information can be propagated across the effective single layer lattice shown in Fig.~\ref{lattice} {\bf (a)}, we consider the prototypical configuration shown in Fig.~\ref{lattice} {\bf (b)}. Here a normalized logical qubit $\ket{Q_{\rm in}}=\mu\ket{0_E}_{1'}+\nu\ket{1_E}_{1'}$ is encoded on the effective qubit $1'$ embodied by the physical qubits $1$ and $2$. After the entangled resource is prepared, the total state of the effective qubits $1'$ and $2'$ 
is written as 
\begin{equation}
\label{DFSstates}
\ket{\phi}_{\rm DFS}=\mu\ket{0_E,+_E}_{1'2'}+\nu\ket{1_E,-_E}_{1'2'}
\end{equation}
with $\ket{\pm_E}=(1/\sqrt 2)(\ket{0_E}\pm\ket{1_E})$. There are two ways to propagate information across effective sub-clusters such as the one considered here. Depending on the physical setup, one strategy may be preferable to the other. The first is to perform a joint measurement on a pair of qubits comprising an effective qubit $i'$ in the basis $B_{i'}(\alpha)=\{\ket{\psi^{+\alpha}}_{i'},\ket{\psi^{-\alpha}}_{i'}\}$ with outcomes $s^{\alpha}_{i'}=\{0,1\}$ and $\ket{\psi^{\pm \alpha}}_{i'}=(1/\sqrt{2})(\ket{01}\pm e^{i \alpha}\ket{10})_{i'_1{i}'_2}$.  
In the case 
of $i'=1'$ in Eq. (\ref{DFSstates}) this strategy simulates the transformation $\sigma_{x}^{s_{1'}^{\alpha} \oplus 1}H_{}R_{z}^{-\alpha}$ on the logical qubit $\ket{Q_{\rm in}}$, where $R_{z}^{-\alpha}$ is a single{\bf -}qubit $z$-rotation in the Bloch sphere by an angle $-\alpha$. 

The second method is to perform single-qubit measurements on $i'_{1}$ and $i'_{2}$ in the bases $B_{i'_1}(\alpha)=\{ \ket{+\alpha}_{i'_1},\ket{-\alpha}_{i'_1}\}$ and $B_{i'_2}(0)=\{ \ket{+}_{i'_2},\ket{-}_{i'_2}\}$ with outcomes $s^{\alpha}_{i'_{1,2}}=\{0,1\}$ and $\ket{\pm \alpha}_{i'_j}=(1/\sqrt{2})(\ket{0}\pm e^{i \alpha}\ket{1})_{i'_j}$ ($j=1,2$). 
For $i'=1'$, $i_{j}'=j$, this simulates the transformation $\sigma_x^{s_{i'_1}^{\alpha} \oplus s_{i'_2}^0 \oplus 1}H R_z^{-\alpha}$ on the logical qubit. 

Consider now the situation depicted in Fig.~\ref{lattice} (c) where we have input logical qubits $\ket{Q^1_{\rm in}}$ and $\ket{Q^2_{\rm in}}$. If no measurements take place on the qubit pairs and the two-qubit gate $S^{13}$ is applied to the top-layer physical qubits $1$ and $3$, we obtain a state that simulates the outcome of the effective gate $S^{1'2'}$ being applied to the logical qubits $1'$ and $2'$. 
These two examples represent the DFS-encoded version of the basic building blocks BBB1 and BBB2 described in~\cite{Tame1}. From the above discussions, one can clearly see how similar the simulations on encoded cluster states are to the original one-way model~\cite{RBH}. In fact with the addition of a third building block, acting on an effective three-qubit cluster structure (whose construction and demonstration in a DFS-encoded scenario goes along the lines depicted above for BBB1 and BBB2) in a straightforward manner), the same concatenation rules described in~\cite{Tame1} can be applied here. The concatenation of the three BBB's is sufficient to simulate any computational process. 

A stabilizer-based approach is also possible in this model by using the correlation relations $G^{(a')}\ket{\phi_{}}_{\cal C'}=(-1)^{\kappa_{a'}}\ket{\phi_{}}_{\cal C'}$ where 
\begin{equation}
\begin{aligned}
G^{(a')}&=X_{a'}\bigotimes_{c'\in {\rm nghb(a')\cap {\cal C'}}} Z_{c'},\\ X_{a'}&=(\sigma_z\sigma_x)_{a_1'}\otimes(\sigma_z\sigma_x)_{a_2'},\\ 
Z_{c'}&=\sigma_{z,c_1'}\otimes \openone_{c_2'}. 
\end{aligned}
\end{equation}
With these tools, one can manipulate the relevant eigenvalue equations defining the cluster resource and design the correct measurement pattern for any unitary simulation~\cite{RBH}. 

All the computational steps can be performed within the DFS and at no point during the computation is the effective cluster state exposed to phase damping type decoherence. In the case of an ideal cluster-resource being produced, this allows the noise effects to be cancelled exactly. However in a real experiment, due to imperfections at the cluster generation stages, we only obtain a state having non-unit overlap with the ideal resource $\ket{\phi}_{\cal C'}$. This results in an effective resource that is partially residing outside the DFS and it is only this fraction that is prone to environmental effects. The benefits of this proposal should now be clear: Encoding in a protected DFS provides us with a method of reducing greatly decoherence processes (ideally, their complete cancellation) in such a way that avoids the use of {\it a posteriori} procedures for correcting the resulting errors.

\subsection{Noise-resilience characterization}

Here we provide a general operative way to determine the effectiveness of the noise protection provided by the realization of one-way QC within a DFS. This can be efficiently done by means of a characterization of the effective map the logical state of a register undergoes in the presence of a noisy computational process. This characterization requires the use of {\it quantum process tomography}~\cite{nielsenchuang}, whose main features we outline next.

A dynamical map ${\cal E}$, that we shall call a ``channel'', acting on the density matrix of a quantum system $\varrho$ is fully identified by the set of Kraus operators $\{\hat{K}_i\}$ such that
\begin{equation}
\varrho\rightarrow{\cal E}(\varrho)=\sum_i\hat{K}_i\varrho \hat{K}^\dag_i,
\end{equation}
with $\sum_i\hat{K}_i^{\dag}\hat{K}_i= \openone$. Channel characterization then reduces to the determination of the $\hat{K}_i$'s. By choosing a complete set of orthogonal operators $\{\hat{\cal K}_m\}$ over which we expand $\hat{K}_i=\sum_m{e}_{im}\hat{\cal K}_m$ we have
\begin{equation}
{\cal E}(\varrho)=\sum_{m,n}\chi_{mn}\hat{\cal K}_m\varrho\hat{\cal K}^\dag_n\end{equation}
with the {\it channel matrix} $\chi_{mn}=\sum_ie_{im}e^{*}_{in}$. This is a pragmatically very useful result as it shows that it is sufficient to consider a fixed set of operators $\{\hat{\cal K}_m\}$, whose knowledge is enough to characterize a channel through the matrix $\chi$. Thus, its matrix elements need to be found. In order to provide them, it is important to notice that the action of the channel over a generic element $\ket{n}\bra{m}$ of a basis in the space of the $d\times{d}$ matrices (and thus $n,m=0,..,d^2-1$), given by ${\cal E}(\ket{n}\bra{m})$, can be determined from a knowledge of the map ${\cal E}$ on the fixed set of states $\ket{n},\,\ket{m},\,\ket{+}=(1/\sqrt 2)(\ket{n}+\ket{m})$ and $\ket{+_y}=(1/\sqrt 2)(\ket{n}+i\ket{m})$ as follows
\begin{equation}
\begin{aligned}
{\cal E}(\ket{n}\bra{m})&={\cal E}(\ket{+}\bra{+})+i{\cal E}(\ket{+_y}\bra{+_y})\\
&-\frac{i+1}{2}[{\cal E}(\ket{n}\bra{n})+{\cal E}(\ket{m}\bra{m})].
\end{aligned}
\end{equation}
Therefore, the effect of the channel ${\cal E}$ on each $\varrho_j=\ket{n}\bra{m}$ (with $j=1,..,d^2$) can be found completely via state tomography of just four fixed states. It is clear that ${\cal E}(\varrho_j)=\sum_k\lambda_{jk}\varrho_k$ as $\{\varrho_k\}$ form a basis, therefore from the above discussion
\begin{equation}
\label{determino}
\begin{aligned}
{\cal E}(\varrho_j)&=\sum_{m,n}\hat{\cal K}_m\varrho_j\hat{\cal K}^{\dag}_n\chi_{mn}=\sum_{m,n,k}\beta^{mn}_{jk}\varrho_k\chi_{mn}\\
& \equiv \sum_k\lambda_{jk}\varrho_k,
\end{aligned}
\end{equation}
where we have defined $\hat{\cal K}_m\varrho_j\hat{\cal K}^\dag_n=\sum_k\beta^{mn}_{jk}\varrho_k$. Therefore we can write
\begin{equation}
\label{beta}
{\lambda_{jk}=\sum_{m,n}\beta^{mn}_{jk}\chi_{mn}}.
\end{equation}
The complex tensor $\beta^{mn}_{jk}$ is set once we make a choice for $\{\hat{{\cal K}}_i\}$ and the $\lambda_{jk}$'s are determined from a knowledge of ${\cal E}(\varrho_j)$. By inverting Eq.~(\ref{beta}), we can determine the channel matrix $\chi$ completely and characterize the map. Let $\hat{U}^\dag$ be the operator diagonalizing the channel matrix (which is always possible for a generic complex matrix that is not a null set with respect to the Lebesgue measure). Then it is straightforward to prove that if $D_i$ are the elements of the diagonal matrix $\hat{U}^\dag\chi{\hat{U}}$, then $e_{im}=\sqrt{D_i}\hat{U}_{mi}$ so that 
\begin{equation}
\hat{K}_{i}=\sqrt{D_i}\sum_j\hat{U}_{ji}\hat{\cal K}_j.
\end{equation}

Important information can be extracted from this characterization for the case of a channel describing a logical qubit transferred across a linear cluster. In particular, we can infer how close a logical output state ${\cal E}(\varrho)$ will be on average to a logical output qubit $\rho$ when no noise is present. Let us label the Schmidt-decomposed bipartite Bell state $\ket{\phi^{+}}$ as $\ket{b}=(1/\sqrt d)\sum_i\ket{i}\ket{i}$ and consider the {\it entanglement fidelity} of the characterized channel~\cite{schumacker} 
\begin{equation}
F_e({\cal E})=\bra{b}(\openone\otimes{\cal E})(\ket{b}\bra{b})\ket{b}.
\end{equation}
This quantifies the resilience of a maximally entangled state to a unilateral action of the channel. $F_{e}(\cal E)$ can easily be determined from the knowledge of the set $\{\hat{K}_i\}$.  
By using the channel entanglement fidelity, the average state fidelity resulting from the application of ${\cal E}$ can be determined as~\cite{nonnoti} 
\begin{equation}
\bar{F}=\frac{1}{3}(2F_{e}({\cal E})+1).
\end{equation}
The theory of quantum process tomography can be applied to the specific experimental setup used for the implementation of DFS encoded one-way QC. The setup dependence is the method used for the state tomography required in order to find the set of output states ${\cal E}(\ketbra{n}{n}),\,{\cal E}(\ketbra{m}{m}),\,{\cal E}(\ketbra{+}{+}),\,{\cal E}(\ketbra{+_y}{+_y})$~\cite{imoto}. Here, we concentrate on a realization in a condensed-matter system where these four state tomographies can be determined through photon-scattering out of an optical lattice embodying the physical support for the entangled resource. However, the technique is easily adapted to any other choice.

\subsection{An application: Information transfer through a linear cluster state}

We now provide an example application of quantum process tomography to a case of interest for our discussion. We concentrate on information flow across both a DFS and standard encoded linear cluster state of three effective qubits. Here, in the standard encoded case, the effective qubits correspond to the physical ones. We assume that each qubit (pair of qubits) in the standard (DFS-encoded) cluster is affected by a phase damping (collective phase damping) decoherence channel characterized by a strength $\Gamma$ that, for the sake of simplicity, we assume to be same for the entire qubit register. The parameter $\Gamma$ can be thought of physically as the rate of damping, or random scattering per unit time of the environment with the qubit systems. In Eq. (\ref{hamil}) this can be taken as the coupling strength of the environment to the qubit-pair system in the final term.

\begin{figure}[t]
\centerline{\psfig{figure=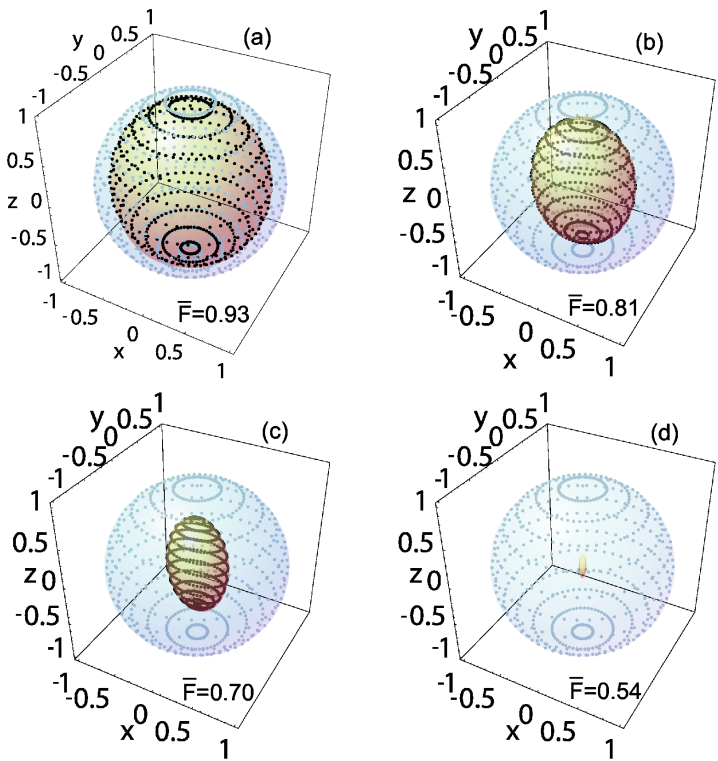,width=9.5cm}}
\caption{Comparison between the DFS and standard evolution of a pure input state transferred across a three element cluster state with phase damping affecting the individual physical qubits. From ${\bf (a)}$ to ${\bf (d)}$, $\Gamma{t}$ is taken to be $0.15,\,0.5,\,1$ and $5$ respectively. The outer blue ball shows the Bloch sphere of output logical qubits in the DFS case. The states are kept pure all along the evolution. The inner ball corresponds to the standard case. An anisotropic shrinking of the Bloch sphere, increasing with $\Gamma{t}$, occurs in a way that quickly decoheres the output states into a totally mixed state. Each dot on a sphere represents a physical density matrix associated with a chosen set $(\theta,\phi)$ for the input state.}
\label{spheres}
\end{figure}

We aim to transfer a quantum state from the first to the last effective qubit in a chain of three elements, which from now on we label $j=1',2',3'$. In the standard one-way model, this implies the measurement of qubits $1'$ and $2'$ in the $B_{1'}(0)$ and $B_{2'}(0)$ bases. In order to fix the ideas, in what follows we concentrate on the case where the measurements have outcomes $s^0_{1'}=s^0_{2'}=0$. This corresponds to the identity operation being carried out on a logical input state. From the discussion in Sec.~\ref{model}, it is clear that the DFS encoding leaves the input state $\ket{in}=\cos{\theta}\ket{0}+e^{i\phi}\cos{\theta}\ket{1}$ unaffected by the noise during the transfer across the chain. On the other hand, a detailed calculation~\cite{qudit} reveals that in the standard encoded case, the state of the logical output qubit residing on qubit $3'$, {\it i.e.} after the performance of the protocol, in the presence of the phase damping environment characterized by the strength $\Gamma$, is written as
\begin{equation}
\label{channelevolution}
\begin{aligned}
\rho_{3'}&=\frac{e^{-\frac{3\Gamma{t}}{2}-i\phi}}{2}(e^{\Gamma{t}/2}\cos\phi-i\sin\phi)\sin(2\theta)\ket{0}_{3'}\!\bra{1}+h.c.\\
&+\frac{1}{2}\openone_{3'}+\frac{e^{-\Gamma{t}/2}\cos(2\theta)}{2}\sigma_{z,3'}.
\end{aligned}
\end{equation}
Having this output state of the effective map undergone by the input logical state $\ket{in}$ and using quantum process tomography, it is possible to compute the corresponding Kraus operators for the logical channel using $\ket{in} \in \{ \ket{0}, \ket{1},\ket{+}, \ket{+_y}\}$, giving
\begin{equation}
\begin{aligned}
\hat{K}_{1}&=e^{-3\tau/8}\sqrt{\sinh\frac{\tau}{4}\cosh\frac{\tau}{2}}\sigma_{x},\\
\hat{K}_{2}&=e^{-3\tau/8}\sqrt{\cosh\frac{\tau}{4}\cosh\frac{\tau}{2}}\openone_{},\\
\hat{K}_{3}&=e^{-3\tau/8}{\cosh\frac{\tau}{4}}\sqrt{2\sinh\frac{\tau}{4}}\sigma_{z},\\
\hat{K}_{4}&=-ie^{-3\tau/8}{\sinh\frac{\tau}{4}}\sqrt{2\cosh\frac{\tau}{4}}\sigma_{y}.
\end{aligned}
\end{equation}
In these equations we took $\tau=\Gamma{t}$. It is straightforward to check that $\sum_{i}\hat{K}^\dag_{i}\hat{K}_i=\openone$ and that $\sum_i\hat{K}_{i}\ket{in}\!\bra{in}\hat{K}^\dag_{i}=\rho_{3'}$. The evolution induced by the Kraus operators associated with the channel, both in the DFS and standard case, is pictorially shown in Fig.~\ref{spheres}. A striking shielding of the quantum information from the action of the environment is revealed. While the standard evolution quickly collapses the state of the output qubit into a maximally mixed state $\openone/2$, the DFS encoded state is kept pure throughout the dynamics and for any value of the decoherence parameter $\Gamma$. In order to provide a full characterization of the channel, in panels ${\bf (a)}$ to ${\bf (d)}$ we give the average state fidelity associated with each instance of the non-DFS channel.
\section{Realization in optical lattices} 
\label{realization}

The effective two-dimensional cluster state shown in Fig.~\ref{lattice} {\bf (a)} can be realized by using alkali-metal atoms such as $^{87}$Rb trapped in a cubic three-dimensional optical lattice. The lattice configuration is achieved with three slightly detuned pairs of counter-propagating laser beams $L^X,L^Y$ and $L^Z$, tuned between the $D1$ and $D2$ line with wavelength $\lambda=785~$nm. The pairs propagate along $\hat{\bf x}$, $\hat{\bf y}$ and $\hat{\bf z}$ respectively and are in a lin$\angle$lin configuration (linearly polarized with electric fields forming an angle $2\theta_{i},~i\in\{x,y,z\}$~\cite{Cal1}), providing lattice sites with periodicity $\lambda/2$ for $\theta_{i}=0,~\forall i$. We assume the lattice is initially loaded with one atom per site, which can be achieved by making a Bose-Einstein condensate undergo a superfluid to Mott insulator (MI) phase transition~\cite{Jak1, Mandel1, W}. Each physical qubit at a lattice site can then be embodied by the single-atom hyperfine states $\ket{h_0}=\ket{0}\equiv\ket{F=2, m_f=2}$ and $\ket{h_1}=\ket{1}\equiv\ket{F=1, m_f=1}$ with $F$ and $m_{f}$ the total angular momentum of the atom and its projection along ${\bf \hat{z}}$ respectively. These states can be coupled via a Raman transition~\cite{Jak1}, using an excited state $\ket{h_e}$ embodied by an additional hyperfine state. Cold controlled collisions using moving trapping potentials between adjacent atoms along the three spatial dimensions can be achieved by individually changing the angles $\theta_{i}$~\cite{Jak2, Mandel1}. The controlled collisions result in a dynamical phase shift applied to only one of the joint states of adjacent atoms, $\ket{0}_a\ket{1}_{a+1} \to -\ket{0}_a\ket{1}_{a+1}$. The entangling operation $\tilde{S}^{ac}=|0 \rangle_a \langle 0| \otimes \sigma_{z,c} + |1 \rangle_a \langle 1| \otimes \openone_c$ is produced and describes a conditional phase shift equivalent to $S^{ac}$, with a $\sigma_z$ operation on qubit $c=(a+1)$. 
\begin{figure*}[t]
\centerline{\psfig{figure=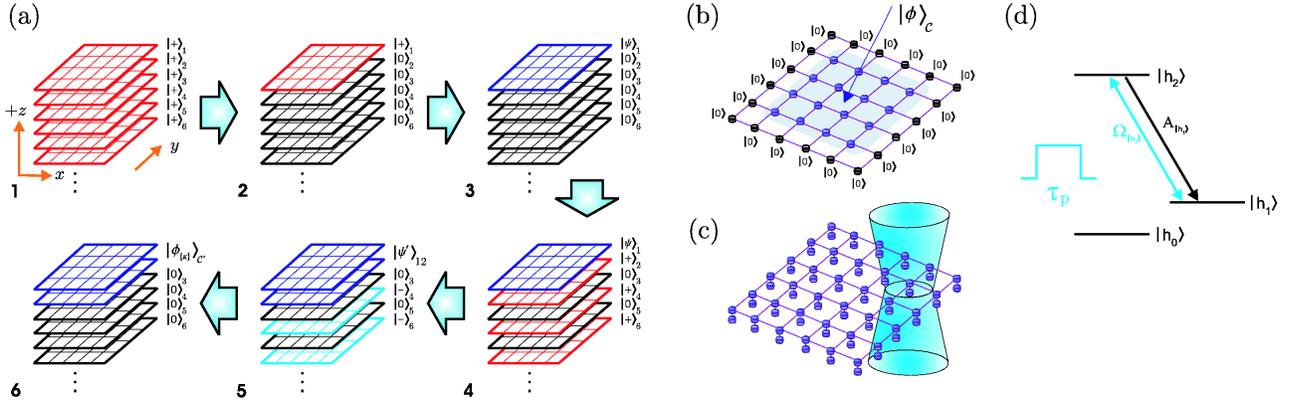,width=17cm}}
\caption{{\bf (a)}: Steps taken to create the effective cluster state in an optical lattice setup. {\bf (b)}: A barrier technique~\cite{adj} to create a cluster state $\ket{\phi_{}}_{\cal C'}$ with the set $\kappa_a=0,~\forall a$ in the central region. {\bf (c)}: Tightly focused laser beam with Gaussian profile used for measurements of the atomic states via fluorescence. {\bf (d)}: Level structure for projective measurements via fluorescence.}
\label{preptop}
\end{figure*}

The initialization of the qubit register prior to any entanglement generation can be achieved by applying Raman transitions to all lattice sites. These can be activated by standing-waves of period $\lambda / 2$ from two pairs of lasers $L_1$ and $L_2$, far blue-detuned by an amount $\Delta$ from the transition $\ket{\{h_0,h_1\}}\leftrightarrow\ket{h_e}$ and oriented along ${\bf \hat{z}}$. All the sites will be located at the maximum-intensity peaks~\cite{Pachos4} and with the atoms initially in $\ket{h_0}$, a rotation of the qubits into the state $\ket{+}$ can be achieved as shown in Fig.~\ref{preptop} {\bf (a)}, step 1. Next, a one-off setting of atoms on all layers (apart from the top layer) to the state $\ket{0}$ can be achieved by either a blurred addressing technique~\cite{Tame3}, interference methods \cite{Joo} or microwave addressing~\cite{Zhang}, see Fig.~\ref{preptop} {\bf (a)}, step 2. To generate entanglement on the top layer only, angles $\theta_x$ and $\theta_y$ are varied so as to apply the operation $\tilde{\cal S}_{=}^{\cal C}=\prod_{a,c\in {\cal C}|a,c\in \gamma_{=}}\tilde{S}^{ac}$. This creates a cluster state with a particular set of eigenvalues $\{\kappa \}$. The non-zero values in this set can be accommodated by modifying the measurement pattern later, as they will determine corresponding values in the final effective lattice \cite{adj}. We denote the resulting cluster state on the top layer as $\ket{\psi}$ as shown in Fig.~\ref{preptop} {\bf (a)}, step 3. It is possible to form standing waves of period larger than $\lambda /2$ with the two pairs of lasers $L_{1,2}$ used for the Raman transitions~\cite{Peil}. Here, the two lasers in each pair are set at angles $\pm\theta/2$ to a given direction $\vec{v}$ on the $x$-$z$ plane. This produces an intensity pattern in the direction perpendicular to $\vec{v}$ on the $x$-$z$ plane with period $d=\lambda /[2 \sin(\theta/2)]$. Thus we can rotate states on the even labeled layers to $\ket{+}$ via a Hadamard rotation $H$, as shown in Fig.~\ref{preptop} {\bf (a)}, step 4. Next, controlled collisions can be initiated along $\hat{\bf z}$ by varying the angle $\theta_z$. This applies the operation $\tilde{\cal S}^{\cal C}_{\|}=\prod_{a,b\in {\cal C}|a,b\in \gamma_{\|}}\tilde{S}^{ab}$ creating an entangled state $\ket{\psi'}$ on the top two layers. Due to the transformation $\ket{0}_a\ket{+}_{b}\to \ket{0}_a\ket{-}_{b}$ from the controlled collisions of the atoms on odd layers $a$ with those on even layers $b$, we produce the structure shown in Fig.~\ref{preptop} {\bf (a)}, step 5. In step 6, we apply the rotation $\tilde{H}:=H\sigma_x$ to all even labeled layers using Raman transitions. 
Finally, the $x$-$y$ lattice spacing is increased from $\lambda/2$ by adiabatically turning on a periodic potential with a larger lattice spacing \cite{Peil}, while turning off the original laser pairs $L^X$ and $L^Y$ \cite{newlatt}.

In order to understand how the DFS encoded cluster state $\ket{\phi_{}}_{\cal C'}$ is generated by the previous steps on the top two layers of the lattice, one needs to consider the operations performed in each step. First, we start with the state $\otimes_{a,b \in {\cal C}}\ket{+,0}_{ab}$ in step 2. Then we apply $\tilde{\cal S}^{\cal C}_{=}$ on the top layer, followed by $H$ to the bottom layer. Finally $\tilde{\cal S}^{\cal C}_{\|}$ is applied between the top and bottom layers and $\tilde{H}$ to the bottom layer. The entire process is
\begin{equation}
\begin{split}
&\prod_{a,b\in {\cal C}|a,b\in \gamma_{\|}}(\openone_a \otimes \tilde{H}_b)\tilde{S}^{ab}(\openone_a \otimes H_b)\times\\
&\prod_{a,c\in {\cal C}|a,c\in \gamma_{=}}\tilde{S}^{ac} \bigotimes_{a,b \in {\cal C}|a,b \in \gamma_{\|}}\ket{+,0}_{ab} 
\end{split}
\end{equation}
which can be reordered to give
\begin{equation}
\begin{split}
\prod_{a,c\in {\cal C}|a,c\in \gamma_{=}}\tilde{S}^{ac}\left[\prod_{a,b\in {\cal C}|a,b\in \gamma_{\|}}(\openone_a \otimes \tilde{H}_b)\tilde{S}^{ab} \bigotimes_{a,b \in {\cal C}}\ket{+,+}_{ab}\right]. 
\end{split}
\end{equation}
The square bracketed part is equivalent to $\otimes_{a,b\in {\cal C}|a,b\in \gamma_{\|}}\ket{\psi^-}_{ab}$. Using a barrier method on the top layer~ \cite{adj} allows $\prod_{a,c\in {\cal C}|a,c\in \gamma_{=}}\tilde{S}^{ac}$ to be formally equivalent to $\prod_{a,c\in {\cal C}|a,c\in \gamma_{=}}S^{ac}$ for the central section of atoms. If this method is not used, then a different set $\{ \kappa \}$ must be taken into account for the effective cluster state in the measurement pattern design. Comparing the above steps with those described in Section~\ref{model}, one can easily see that they create the required effective cluster state $\ket{\phi_{}}_{\cal C'}$. An alternative method for setting up the required effective lattice could be the use of a pattern-formation technique~\cite{Vala} to separate two layers of a three-dimensional lattice from the rest by a gap of at least two layers. As the entanglement is generated via controlled collisions, only the two separated layers will take part in the effective cluster state generation.  
The benefit of the method outlined here is that all other layers are in the state $\ket{0}$, which is important for the measurement stage we are going to discuss. 
\begin{table}[b]
\begin{ruledtabular}
\begin{tabular}{|c|c|c|c|}\hline
Pair $1'$ & Pair $2'$ & Logical & $U_{\Sigma}$ \Large{\phantom{A}} \\ \hline\hline
$\ket{00}_{12}$ & $\frac{1}{\sqrt{2}}[(\mu-\nu)\ket{01}-(\mu+\nu)\ket{10}]_{34}$ & $\sigma_x H \ket{\psi}$ & $\sigma_x$ \Large{\phantom{A}}\\ \hline
$\ket{01}_{12}$ & $-\frac{1}{\sqrt{2}}[(\mu+\nu)\ket{01}-(\mu-\nu)\ket{10}]_{34}$ & $H \ket{\psi}$ & $\openone$ \Large{\phantom{A}}\\ \hline
$\ket{10}_{12}$ & $\frac{1}{\sqrt{2}}[(\mu+\nu)\ket{01}-(\mu-\nu)\ket{10}]_{34}$ & $H \ket{\psi}$ & $\openone$ \Large{\phantom{A}}\\ \hline
$\ket{11}_{12}$ & $-\frac{1}{\sqrt{2}}[(\mu-\nu)\ket{01}-(\mu+\nu)\ket{10}]_{34}$ & $\sigma_x H \ket{\psi}$ & $\sigma_x$ \Large{\phantom{A}}\\ \hline
\end{tabular}
\end{ruledtabular}
\caption{Outcomes from a laser measurement of qubits 1 and 2 in Fig.~\ref{lattice} {\bf (b)}.}
\label{tab1}
\end{table}
\begin{figure*}[t]
\centerline{\psfig{figure=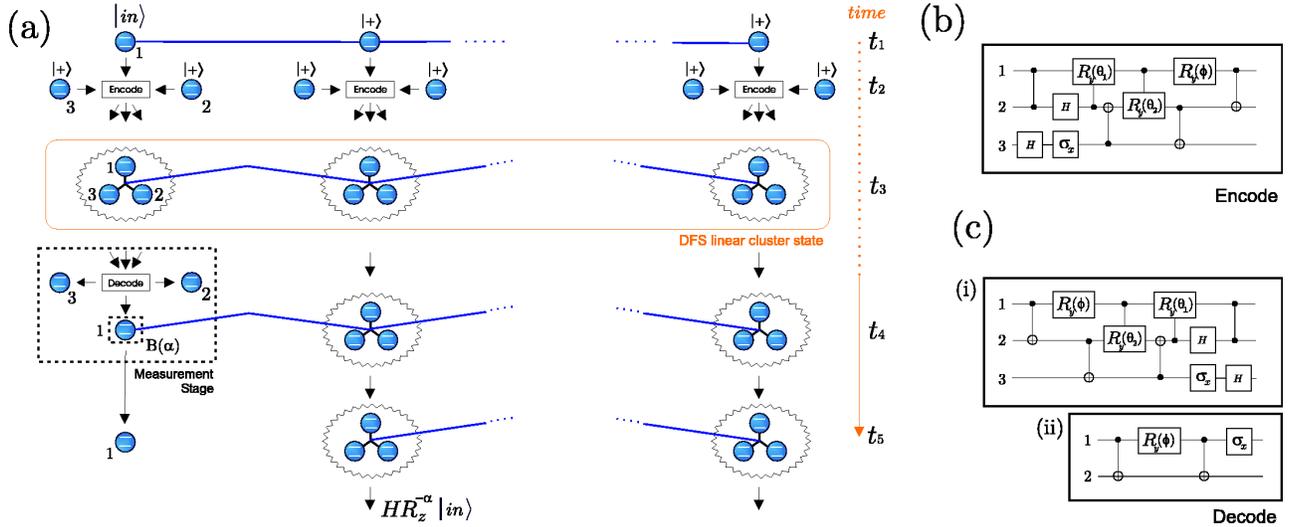,width=17cm}}
\caption{DFS linear cluster state protected from all system-environment coupling terms of the form given in Eq. (\ref{hamil}). {\bf (a)}: Sequence of operations for transferring an arbitrary qubit input state $\ket{in}$. First, the standard cluster state is prepared, then the qubits are encoded (see {\bf (b)}). The only time at which the cluster state is not protected is when the measurements are performed. However, if the measurement stages (which include the decoding stage {\bf (c)}) are carried out in negligible time (with respect to the rate of decoherence), then the remaining cluster after each measurement is never exposed to the environment. {\bf (b):} Encoding Stage, where $\phi=3 \pi/4$, $\theta_1=-\cos^{-1}(\sqrt{2/3})$ and $\theta_2=-\pi/4$. {\bf (c):} Decoding Stage, where $\phi=- 3 \pi/4$, $\theta_1=\cos^{-1}(\sqrt{2/3})$ and $\theta_2=\pi/4$. In (i), qubit 3 is measured in the $\{ \ket{0},\ket{1}\}$ basis and if $\ket{0}_3$ is obtained, then the circuit (ii) must be performed. Qubits 2 and 3 can be discarded after the decoding stage.}
\label{3DFS}
\end{figure*}

In order to perform the measurements, we assume that the $x$-$y$ plane has been expanded such that a single two-qubit pair can be addressed individually in a top-down fashion by a tightly focused laser beam with a Gaussian profile, as schematically shown in Fig.~\ref{preptop} {\bf (c)}~\cite{option1}. This laser is tuned to a hyperfine transition $\ket{h_1}\to \ket{h_2}$ and applied for a pulse-time $\tau_p$ (see Fig.~\ref{preptop} {\bf (d)}). The state $\ket{h_2}$ is taken to have a large spontaneous-emission rate $A_{h_2}$ such that, within the time $\tau_p$, many cycles of absorption-emission will occur (i.e. $A_{h_2}^{-1}\ll\tau_p$). We call $N_{\ket{h_1}}$ the number of photons emitted by a single atom during $\tau_p$ when it is in the state $\ket{h_1}$ and $\eta'$ the ratio of the number of detected photons to emitted photons, due to non-ideal quantum efficiency of the detectors that collect the scattered photons. Starting with the atom in the state $\ket{\psi}=\mu\ket{h_0}+\nu\ket{h_1}$, if one or more photons are detected, the state of the atom is inferred to be $\ket{h_1}$. On the other hand if no photons are detected, the state of the atom is $|\mu|^2 \ketbra{h_0}{h_0}+|\nu|^2 P^d_0\ketbra{h_1}{h_1}$, where $e^{-(1+\eta'/2)\eta'N_{\ket{h_1}}}\leq P^d_0 \leq (1+2\eta'/3)e^{-\eta'N_{\ket{h_1}}}$~\cite{Beige}. Taking $\ket{h_2}$ from the P$_{3/2}$ manifold with $A_{h_2}^{-1}=2.62\times10^{-8}$ \cite{Volz} and a pulse time $\tau_p=2.62\times10^{-6}$, with $\eta'=0.89$~\cite{Rosen}, we can effectively set $P^d_0 = 0$.

Let us consider the measurement laser addressing the two atoms embodying qubits $1$ and $2$ (effective qubit $1'$) as shown in Fig.~\ref{lattice} {\bf (b)} in the top-down fashion described above. Before the laser is applied, we take an encoded state $\ket{\psi}=\mu\ket{0}+\nu\ket{1}$ as being prepared on the first effective qubit and assume a pair of tightly focused lasers $L_1$ and $L_2$ with Gaussian profiles address the lattice along the $x$ and $y$ axes respectively between the top two layers causing the states of qubits $1$ and $2$ to be subject to the Hadamard gate $H$ via a Raman transition. More formally, the operation $H_1\otimes H_2 \otimes \openone_3 \otimes \openone_4$ is applied to the qubits. 
This produces the state 
\begin{equation}
\begin{aligned}
\ket{\phi}_{1'2'}&=(\mu\ket{+}\ket{-}\ket{01}-\mu\ket{+}\ket{-}\ket{10}\\
&-\nu\ket{-}\ket{+}\ket{01}-\nu\ket{-}\ket{+}\ket{10})_{1234}.
\end{aligned}
\end{equation}
The measurement laser is then applied to qubits 1 and 2 projecting the atomic states into the $\sigma_z$ eigenbasis via the fluorescence technique described above. Together with the Hadamard rotations, this carries out a $\sigma_x$ projective measurement.  
A degeneracy in the outcomes exists because both the states $\ket{01}$ and $\ket{10}$ will produce the same statistics of detected photons. However as it can be seen in Table~\ref{tab1}, they apply the same rotations to the logical state upon propagation across to effective qubit $2'$. The byproduct operator $U_{\Sigma}$ which is used to cancel the probabilistic nature of state transfer in one-way QC can therefore be found from a photon-number-resolving detector~\cite{Rosen}. 

In order to carry out an arbitrary measurement along the equatorial plane of the Bloch sphere, one must implement an additional Raman transition prior to the Hadamard rotations. This transition uses a tightly focused laser beam $L_1$ in a top-down fashion along the $z$ axes addressing qubits 1 and 2 and all the qubits below them in that column. This laser together with a paired laser field $L_2$, which has intensity maxima on every odd layer, rotates qubit 1 and all qubits below it on odd layers by $R_z^\alpha$. However, qubits on odd layers below qubit 1 are unaffected as they are in the state $\ket{0}$. Alternative methods for the above processes could be given by an interference approach~\cite{Joo} or microwave addressing~\cite{Zhang}.

\section{remarks} 
\label{summary}
In this work we have provided a proposal for one-way QC carried out within a DFS of a supporting quantum system. Our model integrates, for the first time, one of the most promising models for QC and an effective strategy for information protection. We have also described a possible optical lattice setup as an example to show how this may be done in a physically realizable setting. The resilience to noise induced by the encoding into a DFS can be quantified by means of quantum process tomography as we have shown. 
So far, only phase damping errors have been considered in our scheme. However it is possible to extend the approach to the construction of a DFS offering protection from all types of environmental error resulting from Eq. (\ref{hamil}). In Fig. \ref{3DFS} we sketch the steps for the achievement of full protection. The scheme is inspired by recent work \cite{Viola} to which we refer for further details. The encoding is given by $\{\ket{0_E}_{1'}:=(1/\sqrt{2})(\ket{10}-\ket{01})_{12}\ket{0}_{3}, \ket{1_E}_{1'}:=(2/\sqrt{6})\ket{0}_1(\ket{10}-\ket{01})_{23}+ (1/\sqrt{6})(\ket{10}-\ket{01})_{12}\ket{0}_3 \}$, where now three entangled physical qubits (instead of two) embody a single effective cluster qubit. An important difference here with respect to the phase damping DFS is that now encoding (see Fig. \ref{3DFS} {\bf (b)}) and decoding stages (see Fig. \ref{3DFS} {\bf (c)}) are essential for providing the protection and recovery of the cluster state. It must be stressed that the description we give here is not the most economical or optimal one. Development of the scheme shown in Fig. \ref{3DFS}, with a minimal resource perspective is needed and is the topic of our current study. This could represent a powerful {and novel} technique for the protection of one-way QC performed in systems exposed to environmental effects. It would also represent an important simplification with respect to current proposals for noise-resilient measurement-based QC. 

\acknowledgments

We acknowledge discussions with R. Prevedel, A. Stefanov, T. Jennewein, D. Feder, M. Garrett and R. Stock. We thank DEL, the Leverhulme Trust (ECF/40157), and the UK EPSRC for financial support.



\begin{thebibliography}{99}
\bibitem{newlatt} This new enlarged lattice spacing is restricted by the validity of the Mott Insulator regime \cite{Jak1}.
\bibitem{option1} Position dependent energy shifts from focused lasers and an addressing microwave laser is also an option \cite{Zhang}.
\end{thebibliography}
\end{document}